\documentclass[aps,prl,floats,showpacs,twocolumn,superscriptaddress]{revtex4}
\usepackage{graphicx,color}

\begin{document}
\title{Zero-bias anomaly in the tunneling density of states of graphene}
\author{Eros Mariani}
\affiliation{Institut f\"ur Theoretische Physik, Freie Universit\"at Berlin, Arnimallee 14, 14195 Berlin, Germany}
\author{Leonid I. Glazman}
\affiliation{W.I.\ Fine Theoretical Physics Institute, University of Minnesota, Minneapolis, MN 55454, USA}
\author{Alex Kamenev}
\affiliation{Department of Physics, University of Minnesota, Minneapolis, MN 55454, USA}
\author{Felix von Oppen}
\affiliation{Institut f\"ur Theoretische Physik, Freie Universit\"at Berlin, Arnimallee 14, 14195 Berlin, Germany}
\date{\today}
\begin{abstract}
In the vicinity of the Fermi energy, the band structure of graphene is well described by a Dirac equation. Impurities will generally induce both a scalar potential as well as a (fictitious) gauge field acting on the Dirac fermions. We show that the  angular dependence of the zero-bias anomaly in the spatially resolved tunneling density of states (TDOS) around a particular impurity allows one to distinguish between these two contributions. Our predictions can be tested in scanning-tunneling-microscopy measurements on graphene. 
\end{abstract}
 \maketitle

{\em Introduction}---Since its recent experimental realization \cite{Geim,Kim}, graphene -- a monolayer of graphite -- has attracted a lot of attention due to its remarkable electronic structure. At low doping, the Fermi surface of graphene lies in the vicinity of two points in the Brillouin zone (termed Dirac points or valleys) near which the spectrum is characterized by a massless Dirac dispersion \cite{Gonzales,Dresselhaus}
with velocity $v$. Pioneering experiments on this novel 2D electron system have shown that the Dirac nature of carriers induces an anomalous integer quantum Hall effect as well as a finite conductivity at vanishing carrier density \cite{Geim,Kim}.

Within the independent-electron approximation, the density of carrier states $\nu(E)=|E|/2\pi \hbar^2 v^2$ vanishes linearly at the Dirac point $E=0$. Thus, Coulomb interactions remain unscreened when the Fermi energy is at the Dirac point, which leads to deviations from conventional Fermi-liquid expectations \cite{Gonzales96}. The Fock diagram for the electronic self energy, shown in Fig.\ 1(a), entails a logarithmic correction to the linear dispersion relation, $E(p)=[v+(e^2/4)\ln(D/vp)]p$, with $D$ the bandwidth, and hence to the tunneling density of states
\begin{equation}
   \nu(E)\simeq\frac{|E|}{2\pi \hbar^2[v+(e^2/4)\ln(D/|E|)]^2}\quad .
\end{equation}
At the same time, there is no renormalization of the electron charge from the diagram in Fig.\ 1(b). 
Several authors have discussed possible instabilities of the electron system in graphene \cite{Herbut,Tsvelik,here}, although at present, there is no related experimental evidence. 

Impurities and defects generally induce both a scalar potential as well as a fictitious gauge field in the effective Dirac description \cite{Ludwig}. Site disorder in the underlying tight-binding model is associated with a random scalar potential. Random hopping due to lattice deformations leads to a random gauge field which is abelian (nonabelian) for intravalley (intervalley) scattering. The presence of a random gauge field leads to rich weak-localization physics \cite{GeimWL,FalkoWL,MorpurgoWL}. Several recent works have addressed  localization for Dirac electrons \cite{AleinerEfetov,Altland,Mirlin}.

Unlike conventional two-dimensional electron systems (2DES), the 2DES in graphene is exposed at the surface, making it directly accessible to local-probe measurements, such as scanning tunneling microscopy (STM) or scanning single-electron transistors \cite{Yacoby}. Motivated by this fact, we investigate the tunneling density of states (TDOS) in this paper. Unlike the logarithmic corrections due to Coulomb interactions which become singular at the Dirac point, the combined effects of disorder and interactions at finite doping lead to a zero-bias anomaly (ZBA) which is tied to the Fermi energy. 

The zero-bias anomaly in disordered conductors arises from scattering of electrons on impurities and on the potential generated around the impurity by the Friedel oscillations \cite{AAL,Rudin}. This combined effect of disorder and interactions leads to a suppression of the tunneling density of states at low biases $\omega$. 
The ZBA in graphene raises several issues. (i) The wavelength of Friedel oscillations in graphene \cite{Cheianov} diverges as the Fermi energy approaches the Dirac point. At the same time, the strength of the Coulomb interaction increases due to the absence of screening at the Dirac point. We study how this competition affects the magnitude of the ZBA as the Fermi energy $E_F$ approaches the Dirac point within the regime $E_{F}^{}\tau >1$ (with $\tau$ the elastic mean free time due to impurity scattering). (ii) We show that the zero-bias anomaly provides a means of distinguishing experimentally between the scalar potential and the fictitious gauge potential induced in the Dirac equation by a specific impurity. Our considerations focus on the quasiballistic regime $E_F>\omega> 1/\tau$ relevant in relatively clean samples. In this regime, the physics is essentially captured by considering electrons scattering off isolated impurities. Our approach breaks down at the lowest biases $\omega<1/\tau$, where electron diffusion becomes relevant \cite{Khveshchenko}.

The use of the Dirac formalism restricts us to a discussion of the local TDOS which is coarse-grained over a region large compared to the lattice spacing. At the same time, the long-wavelength nature of the Dirac description allows us to include the effects of electron-electron interactions on the local TDOS which have been neglected in several earlier studies performed in the framework of a tight-binding model \cite{GuineaTDOS,Balatsky}. As we argue below, the effects of electron-electron interactions on the coarse grained TDOS are of particular importance in graphene since the effects of impurity scattering alone are suppressed by the chiral symmetry of Dirac fermions. 

\begin{figure}
	\centering
		\includegraphics[width=0.8\columnwidth]{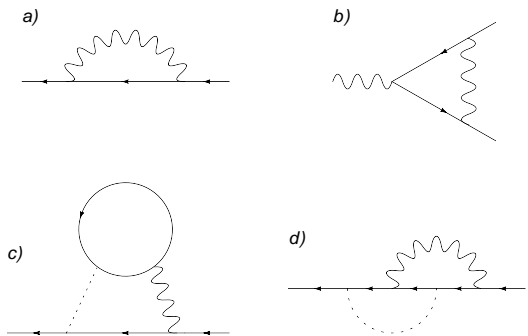}
			\caption{ (a) Fock self energy of Dirac fermions; (b) leading vertex correction; (c) Hartree and (d) Fock diagram for the zero-bias anomaly.
			\label{feynman}}
\end{figure}

{\em Graphene}---Within the tight-binding approximation, the bandstructure of graphene is described by the Hamiltonian  
$H=-t \sum_{\langle ij\rangle}^{}c^{\dagger}_{i}c^{}_{j} + {\rm h.c.}$ on a hexagonal lattice, where $t$ is the hopping matrix element, $c^{}_{i}$ is the annihilation operator of an electron on lattice site $i$, and only nearest-neighbor hopping has been considered. The 2D hexagonal lattice consists of two identical sublattices A and B, and thus two sites per unit cell. We choose the vectors connecting a B site with the neighboring A sites as $\mathbf{e}^{}_{1}=a\, (-1,0)$, $\mathbf{e}^{}_{2}=a\, (1/2,\sqrt{3}/2)$ and $\mathbf{e}^{}_{3}=a\, (1/2,-\sqrt{3}/2)$ with $a$ being the bond length. The Bloch spectrum of the tight-binding Hamiltonian has zero energy (corresponding to the Fermi energy at half filling) at two inequivalent Dirac points in the reciprocal lattice, which we choose to be $\mathbf{k}^{}_{\pm}=\pm\mathbf{k}_{D}^{}$, with $\mathbf{k}_{D}^{}=2\pi/(3\sqrt{3}a)\, (\sqrt{3},1)$. Linearizing the spectrum about both Dirac points and arranging the Hamiltonian in a $4\times 4$ matrix form, one can write
(see e.g.\ Ref.\ \cite{Cheianov})
\begin{equation}
\label{HDirac}
H=\hbar v\,\mathbf{\Sigma}\cdot\mathbf{k}
\end{equation}
with $v=3ta/2$, $\mathbf{k}$ the 2D wavenumber deviation from the Dirac point, $\Sigma_{x,y}^{}=\Pi_{z}^{}\otimes\mathbf{\sigma}_{x,y}^{}$ the components of the vector $\mathbf{\Sigma}$ where $\Pi_{i}^{}$ and $\sigma_{j}^{}$ are Pauli matrices acting in the spaces of the Dirac points $(+/-)$ and the sublattices $(A/B)$, respectively. The Hamiltonian in Eq.\ (\ref{HDirac}) acts on the four-component spinors $(u_{A,\mathbf{k}}^{+},u_{B,\mathbf{k}}^{+},u_{B,\mathbf{k}}^{-},u_{A,\mathbf{k}}^{-})$ of Bloch amplitudes in the $A/B$ and $+/-$ channels.

Within the Dirac Hamiltonian Eq.\ (\ref{HDirac}), an impurity localized, say, at the origin can be quite generally accounted for by adding a general local operator consistent with time-reversal symmetry,  
\begin{equation}
\label{Hdis}
H_{\mathrm{dis}}^{}(\mathbf{r})=\hat{u}\,\delta (\mathbf{r}).
\end{equation}
Here, $\hat{u}$ is a matrix in the $4\times 4$ Dirac space, reflecting inhomogeneities in both site energies and hopping amplitudes of the underlying tight-binding Hamiltonian \cite{AleinerEfetov,Cheianov}. The matrix $\hat u$ can be parametrized in terms of ten real numbers $u_{0}^{},\, u_{sl}^{}\, (s,l=x,y,z)$ as
\begin{equation}
\label{Param}
\hat{u}=u_{0}^{}\,\hat{I}+\sum_{s,l=x,y,z}^{}u_{sl}^{}\Sigma_{s}^{}\Lambda_{l}^{}
\end{equation}
with $\hat{I}=\Pi_{0}^{}\otimes\sigma_{0}^{}$ the identity matrix and $\Sigma_{z}^{}=\Pi_{0}^{}\otimes\sigma_{z}^{}$, $\Lambda_{x,y}^{}=\Pi_{x,y}^{}\otimes\sigma_{z}^{}$, $\Lambda_{z}^{}=\Pi_{z}^{}\otimes\sigma_{0}^{}$.

The parameters $u_0,u_{sl}$ can be classified according to whether they describe intravalley or intervalley scattering
and whether they take the form of potential or (fictitious) gauge field disorder in the Dirac equation. For example, intervalley scattering will involve $\Pi_{x,y}$ and contributions to the fictitious gauge field involve $\Sigma_{x,y}$.
The complete classification is detailed in Table I. 

\begin{table}[htdp]
\begin{center}
\begin{tabular}{|c|c|c|}\hline
{} &  potential & (fictitious) gauge field\\ \hline \hline
			  intra-valley & $\; u_0^{}\, ,\, u_{zz}^{}\;$ & $\; u_{xz}^{}\, ,\, u_{yz}^{}\;$ (abelian) \\  \hline
			  inter-valley & $\; u_{zx}^{}\, ,\, u_{zy}^{}\; $ & $\; u_{xx}^{}\, ,\, u_{xy}^{}\, ,\, u_{yx}^{}\, ,\, u_{yy}^{}\; $ (nonabelian) \\ \hline
\end{tabular}
\end{center}
\label{}
\caption{Physical significance of the various components of the disorder potential $\hat u$  in terms of type of potential (columns) and type of scattering (rows).}
\end{table}

For Dirac fermions described by Eq.\ (\ref{HDirac}), the retarded Green function with energy $\epsilon$ and momentum $\mathbf{p}$ is
$\mathbf{G}^{R}_{\epsilon} (\mathbf{p})=\sum_{\pm}{\hat{s}_{\mp\mathbf{p}}^{}}/[{\epsilon \pm vp+i\eta}]$
where $\hat{s}_{\mathbf{p}}^{}=1/2\left[\mathbf{1}+\mathbf{\Sigma}\cdot\mathbf{p}/p\right]$ denotes the projector onto the state with positive chirality $\mathbf{\Sigma}\cdot\mathbf{p}/p$ and $\eta\rightarrow 0^{+}_{}$. In view of particle-hole symmetry, we specify attention to electron doping (positive Fermi energy). Then, in the regime $k_{F}^{}r\gg 1$, the real space Green function takes the form
\begin{equation}
\label{Green-r}
\mathbf{G}^{R}_{\epsilon} (\mathbf{r},0)\simeq -\frac{e_{}^{i\pi /4}p_{\epsilon}^{}}{\sqrt{2\pi}\hbar v}\frac{e_{}^{ip_{\epsilon}^{}r}}{\sqrt{p_{\epsilon}^{}r}}\left[\hat{s}_{\mathbf{r}}^{}+\frac{i}{4p_{\epsilon}^{}r}\,\mathbf{\Sigma}\cdot\frac{\mathbf{r}}{r}\right]
\end{equation}
with $p_{\epsilon}^{}=\epsilon/\hbar v$. Here, we retained the next-to-leading term in $1/k_Fr$ in the square bracket for later convenience. The projector $\hat{s}_{\mathbf{r}}$ is defined in analogy with $\hat{s}_{\mathbf{p}}$.

{\em Tunneling density of states}---To compute the (coarse-grained) local TDOS 
\begin{equation}
    \nu_\epsilon({\bf r}) =-\frac{2}{\pi}{\rm Tr}\, {\rm Im} {\cal G^R_\epsilon}({\bf r},{\bf r}) 
\label{TDOS_def}
\end{equation}
in the quasiballistic regime, we consider contributions to the full single-particle Green function ${\cal G^R_\epsilon}$ due to interactions and scattering off isolated impurities. The leading-order correction arises from paths involving a single impurity scattering event, corresponding to direct backscattering of electrons to the tunneling point. 

In conventional 2DES, this contribution falls off as $1/r$ and oscillates with wavevector $2k_F$ ($r$ denotes the distance of the tunneling point from the impurity). For Dirac electrons, the leading term is suppressed by chirality and a straight-forward calculation yields a faster spatial decay,
\begin{equation}
   \frac{\delta \nu_{F}^{}({\bf r})}{\nu_{F}^{}} = -4\, \nu_{F}^{}\, u_{0}^{}\, \frac{\sin (2k_{F}^{}r)}{(k_{F}^{}r)^{2}_{}}\quad .
   \label{NuOnlyU}
\end{equation}
Due to this suppression, one expects the combined effects of disorder and interactions, involving scattering on the impurity as well as on the potential generated by the surrounding Friedel oscillations, to be particularly important
in graphene. It is the latter contribution which we now address.

In the presence of a finite Fermi surface, the impurity potential generates Friedel oscillations in the carrier density. This in turn yields an additional oscillatory scattering potential due to the Coulomb interaction, affecting the  electronic return probability and consequently the TDOS. To first order in the electron-electron interaction and the impurity potential, the relevant correction to the Dirac fermion Green function \cite{Rudin}
is given by 
\begin{eqnarray}
\label{DeltaG}
\delta {\cal G}^{R}_{\epsilon} (\mathbf{r},\mathbf{r})=\int d\mathbf{r}^{}_{1}d\mathbf{r}^{}_{2}\, \mathbf{G}^{R}_{\epsilon} (\mathbf{r},0)\hat{u}\mathbf{G}^{R}_{\epsilon} (0,\mathbf{r}_{1}^{}) && \nonumber\\ \times\hat{H}_{\rm HF}^{}(\mathbf{r}_{1}^{},\mathbf{r}_{2}^{})\mathbf{G}^{R}_{\epsilon} (\mathbf{r}_{2}^{},\mathbf{r})
  +({\hat u}\leftrightarrow H_{\rm HF}),&&
\end{eqnarray}
in terms of the Hartree and Fock potentials $\hat{H}_{\rm HF}(\mathbf{r}_{1},\mathbf{r}_{2})=\hat{V}_{H}(\mathbf{r}_{1})\delta (\mathbf{r}_{1}-\mathbf{r}_{2})-\hat{V}_{F}(\mathbf{r}_{1},\mathbf{r}_{2})$ with
$\hat{V}_{H}(\mathbf{r}_{1}) = \mathrm{Tr}[\int d\mathbf{r}^{\prime}V(\mathbf{r}_{1}-\mathbf{r}^{\prime})\delta\rho (\mathbf{r}^{\prime},\mathbf{r}^{\prime})]$ and 
$\hat{V}_{F}^{}(\mathbf{r}_{1}^{},\mathbf{r}_{2}^{})=\frac{1}{2}V(\mathbf{r}_{1}^{}-\mathbf{r}_{2}^{})\delta\rho (\mathbf{r}^{}_{1},\mathbf{r}^{}_{2})$.
Here, $\rho (\mathbf{r},\mathbf{r}^{\prime}_{})=2\sum_{k\le k_{F}^{}}^{}\psi_{\mathbf{k}}^{} (\mathbf{r})\otimes\psi_{\mathbf{k}}^{\dagger}(\mathbf{r}_{}^{\prime})$ denotes the density matrix and $V({\bf r})$ the screened Coulomb interaction. The shorthand $({\hat u}\leftrightarrow H_{\rm HF})$ denotes a similar contribution with the order of ${\hat u}$ and $H_{\rm HF}$ interchanged and appropriate changes to the spatial arguments.
To first order in the impurity potential Eq.\ (\ref{Hdis}), the density matrix emerges from the corrections
\begin{equation}
\label{DeltaPsi}
\delta\psi_{\mathbf{k}}^{}(\mathbf{r})=\int\frac{d\mathbf{p}}{\left(2\pi\right)^{2}_{}}\frac{\hat{s}_{\mathbf{p}}^{}}{E_{k}^{} -vp+i\eta}\, e^{i\mathbf{p}\cdot\mathbf{r}}\,\hat{u}\, |\mathbf{k}\rangle\; .
\end{equation}
to the plane-wave wavefunctions $\psi_{\mathbf{k}}^{}(\mathbf{r})=e^{i\mathbf{k}\cdot\mathbf{r}}|\mathbf{k}\rangle$ with energy $E_{k}^{}$.
The corresponding correction to the density matrix takes the form $\delta\rho (\mathbf{r},\mathbf{r}^{\prime}_{})=R(\mathbf{r},\mathbf{r}_{}^{\prime})+R^{\dagger}_{}(\mathbf{r}_{}^{\prime},\mathbf{r})$ with
\begin{eqnarray}
\label{R}
R(\mathbf{r},\mathbf{r}_{}^{\prime})\simeq \frac{ik_{F}^{}}{8\pi^{2}_{}v\sqrt{rr^{\prime}_{}}}\big[\frac{e^{ik_{F}^{}\left(r-r^{\prime}_{}\right)}_{}}{r-r^{\prime}_{}}\, 4i\hat{s}_{\mathbf{r}}^{}\hat{u}\hat{s}_{\mathbf{r}^{\prime}_{}}^{}+ && \nonumber \\
+\frac{e^{ik_{F}^{}\left(r+r^{\prime}_{}\right)}_{}}{r+r^{\prime}_{}}\, 4\hat{s}_{\mathbf{r}}^{}\hat{u}\hat{s}_{-\mathbf{r}^{\prime}_{}}^{}\big] &&
\end{eqnarray}
in leading order in $1/k_{F}^{}r\ll 1$.

\begin{figure}
	\centering
		\includegraphics[width=0.6\columnwidth]{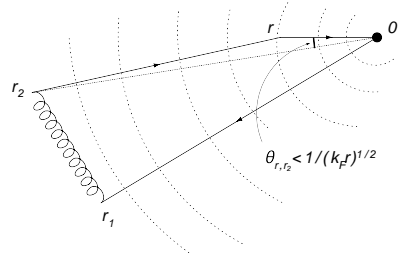}
			\caption{ Schematic of the stationary paths for the Fock contribution  
			\label{FockPaths}}
\end{figure}

It has been shown that due to the chiral symmetry of Dirac fermions, the Friedel oscillations of the electron density decay as $1/r^3$ as opposed to $1/r^2$ in conventional two-dimensional electron systems \cite{Cheianov}. For this reason, the Hartree contribution to the zero-bias anomaly is suppressed and to leading order, we only need to consider the Fock contribution. 
Then, the contribution Eq.\ (\ref{DeltaG}) corresponds to a particle being injected at $\mathbf{r}$, moving to the impurity (located at the origin), being scattered to $\mathbf{r}_{1}^{}$ where it experiences the non-local Fock potential, and finally returning from $\mathbf{r}_{2}^{}$ to the injection point $\mathbf{r}$.
The dominant contribution to the integral can be identified by analyzing the phase factors of the integrand, cf.\ Fig.\ \ref{FockPaths} and Ref.\ \cite{Rudin}. Clearly, paths entailing phases which oscillate rapidly with $\mathbf{r}_{1}^{}$ and $\mathbf{r}_{2}$ lead to negligible corrections.
The three retarded Green functions in Eq.\ (\ref{DeltaG}) yield a phase factor $\exp[i p_{\epsilon}^{}(r+r_{1}^{}+|\mathbf{r}_{2}^{}-\mathbf{r}|)]$ while the density matrix yields $\exp[i\alpha k_{F}^{}(r_{1}^{}+\beta r_{2}^{})]$, with $\alpha ,\beta =\pm 1$. We are interested in corrections to the TDOS in the vicinity of the Fermi energy, so that $\epsilon =E_{F}^{}+\hbar\omega$, with $\hbar\omega\ll E_{F}^{}$ and $p_{\epsilon}^{}=k_{F}^{}+\omega/v$. A correction to the TDOS which varies slowly with $r$ arises from the region $r_{2}^{}\gg r$, so that $|\mathbf{r}_{2}^{}-\mathbf{r}|\simeq r_{2}^{}-r\cos \theta$, where $\theta$ is the angle between $\mathbf{r}$ and $\mathbf{r}_{2}^{}$. Indeed, the oscillatory dependence on $r$ is essentially cancelled if $|\theta |\lesssim 1/\sqrt{p_{\epsilon}^{}r}\simeq 1/\sqrt{k_{F}^{}r}\ll 1$, yielding $\mathbf{r}_{2}^{}\simeq r_{2}^{}\hat{\mathbf{r}}$ in leading order. The remaining phase dependence on $r_{1}^{}$ and $r_{2}^{}$ is $\exp[i p_{\epsilon}^{}(r_{1}^{}+r_{2}^{})]$, which can be cancelled by choosing $\alpha =-1$ and $\beta=1$ in the density matrix, i.e., by the rapidly oscillating term $\exp[-i k_{F}^{}(r_{1}^{}+r_{2}^{})]$ in $R^{\dagger}_{}(\mathbf{r}_{2}^{},\mathbf{r}_{1}^{})$. This cancellation is valid as long as $r_{2}^{}<v/\omega$.
As a result, we obtain 
\begin{equation}
\label{DeltaGFockResult}
\delta {\cal G}^{R}_{\omega} (\mathbf{r},\mathbf{r})\simeq \frac{e_{}^{i\pi /4}\tilde{V}(0)\, k_{F}^{2}}{\sqrt{2\pi}2\pi_{}^{3}\hbar^{4}_{}v^{4}_{}}\frac{1}{r^{2}_{}}\left[\hat{s}_{\mathbf{r}}^{}\hat{u}\hat{s}_{-\mathbf{r}}^{}\hat{u}\hat{s}_{\mathbf{r}}^{}\right]
\end{equation}
with $\tilde{V}(0)$ being the Fourier transform of the screened Coulomb interaction at zero momentum.
The correction to the local TDOS, normalized to the bare DOS of graphene at the Fermi level, $\nu_{F}^{}=k_{F}^{}/(2\pi \hbar v)$, is then 
\begin{equation}
\label{DeltaNuLocal}
\frac{\delta\nu^{}_{\omega} (\mathbf{r})}{\nu_{F}^{}}\simeq -\frac{4 \nu_{F}^{3}\tilde{V}(0)}{\pi_{}^{3/2}}\frac{1}{(k_{F}^{}r)^{2}_{}}\mathrm{Tr}\left[\hat{s}_{\mathbf{r}}^{}\hat{u}\hat{s}_{-\mathbf{r}}^{}\hat{u}\right],
\end{equation}
valid in the regime $1/k_{F}^{}< r < v/\omega$. The correction saturates for $r<1/k_{F}^{}$ and decays as $1/r_{}^{3}$, for $r> v/\omega$. For a finite density $n_i$ of impurities, spatial averaging of Eq.\ (\ref{DeltaNuLocal}) 
(or, alternatively, the diagrams in Figs.\ 1(c) and (d)) yields the average TDOS  
\begin{equation}
\label{DeltaNuAv}
\frac{\delta\nu^{}_{\omega}}{\nu_{F}^{}}\simeq \frac{8\nu_{F}^{2} n_{i}^{}}{\sqrt{\pi}k^{2}_{F}}\mathrm{Tr}\left[2\hat{u}_{}^{2}-\sum_{\alpha =x,y}^{}\hat{\Sigma}_{\alpha}^{}\hat{u}\hat{\Sigma}_{\alpha}^{}\hat{u}\right]\,\ln\left(\frac{\hbar \omega}{E_{F}^{}}\right)
\end{equation}
where we employed the Thomas-Fermi expression $\tilde{V}(0)=1/\nu_{F}^{}$ for the 2D Coulomb interaction.
Similar to conventional two-dimensional electron systems \cite{Rudin}, graphene also exhibits a logarithmic
zero-bias anomaly. Eq.\ (\ref{DeltaNuAv}) shows that the relative strength of the ZBA is only logarithmically dependent on the Fermi energy (since $\nu_F\propto k_F$), even though the wavelength of the Friedel oscillations diverges as $E_F$ approaches the Dirac point. This is a consequence of the reduced screening of the Coulomb interaction.

It is rewarding to also consider the local TDOS, specifically its angular dependence around an impurity. Indeed, different kinds of disorder, as parametrized by the elements $u_{sl}^{}$ in Eq.\ (\ref{Param}), yield different angular dependences: While potential disorder in the Dirac equation yields a local TDOS which is rotationally symmetric around an isolated impurity, gauge-field disorder, be it abelian or nonabelian, leads to a dipole-like angular dependence of the local TDOS. 
Indeed, we find that 
\begin{equation}
\label{DeltaNuzz}
\frac{\delta\nu^{}_{\omega} (\mathbf{r})}{\nu_{F}^{}}\simeq -\frac{8 \nu_{F}^{2}}{\pi_{}^{3/2}}\frac{A+B\sin^{2}_{}\phi+C \cos^{2}_{}\phi}{(k_{F}^{}r)^{2}_{}}.
\end{equation}
with $\phi$ the angle between the vector $\mathbf{r}$ and the direction of the bond with modified tunneling amplitude. Here, $A=u_{zx}^{2}+u_{zy}^{2}+u_{zz}^{2}$ arises from potential scattering, while $B=u_{xz}^{2}+u_{yx}^{2}+u_{yy}^{2}$ and $C=u_{xx}^{2}+u_{xy}^{2}+u_{yz}^{2}$ arise from the gauge field induced by the impurity. Thus, measurements of the local TDOS can be employed to distinguish between the random potential and the random gauge field induced by an impurity in the Dirac equation. 

\begin{figure}[ht]
	\centering
		\includegraphics[width=0.65\columnwidth]{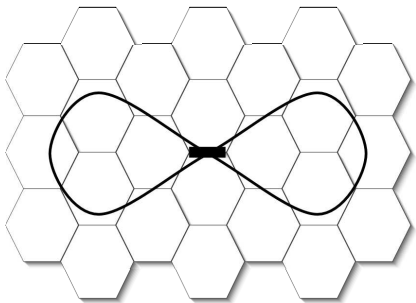}
			\caption{Angular dependence of the zero-bias anomaly in the coarse-grained TDOS around an impurity inducing a fictitious gauge field in the Dirac equation.
			\label{angular}}
\end{figure}

This result is illustrated in Fig.\ \ref{angular}. Impurities induce a gauge field in the Dirac equation if they modify a hopping amplitude. The affected bond then defines the preferred direction of the "dipolar" pattern of the local TDOS.

{\em Conclusions}---A distinctive feature of the two-dimensional electron system in graphene is its exposure at the surface, making it directly accessible to local-probe techniques including STM. We find that such experiments can be employed to gain information about the character of impurities (potential or gauge-field) by  exploring the angular dependence of the local TDOS around a specific impurity. There is no angular dependence in the contribution Eq.\ (\ref{NuOnlyU}) due to disorder scattering alone. Angular dependence of the local TDOS emerges in the zero-bias anomaly due to Coulomb interactions and impurity scattering, and arises from gauge-field disorder only. 

This work was supported in part by DOE Grant DE-FG02-06ER46310 (LIG and
AK), as well as by DIP (FvO). One of us (FvO) enjoyed the hospitality of the TPI at the University of Minnesota, where this work was initiated, and of the KITP Santa Barbara (NSF Grant PHY99-07949) while it was completed.


\begin{thebibliography}{99}
\bibitem{here} In this paper, we assume that a perturbative treatment of the Coulomb interaction is a reasonable starting point, either because these instability scenarios do not apply to graphene or because we consider the system sufficiently far above the corresponding transition temperature.
\end{thebibliography}
\end{document}